# Majorana Representation in Mathematical Modeling of Quantum States

Farhod Shokir

*S.U.Umarov Physical -Technical Institute*
*of the National Academy of Sciences of Tajikistan*

**Abstract:** In this paper, using the Majorana method, mathematical modeling of the state of quantum systems with spin number $S = j\hbar$. An expression for the correlation functions of oriented states in the general case $j \geq \frac{1}{2}$ is obtained.

**Keywords**: Majorana representation, Bloch sphere, mathematical modeling, quantum systems, spin, correlation functions.

## 1. Introduction

The study of quantum systems with a high spin value $|\psi\rangle^{(S>j\hbar)}$ ($j > \frac{1}{2}$)) is of great fundamental importance, including in elementary particle physics, for classifying entanglement in symmetric quantum states, studying high-spin Bose condensate, calculation of geometric phases of high-spin systems, statistics of chaotic quantum-dynamic systems, etc. [1]. However, the direct geometric interpretation of quantum states with a high spin value ($S > \frac{1}{2}\hbar$) is problematic, since it is quite difficult to visualize the processes occurring in the multidimensional space $\mathbb{R}^D$ ($D \geq 4$). In this case, the description of a quantum system with SU(2) symmetry and its evolution using Majorana's method [2] provides us with an intuitive way to study systems with $S > \frac{1}{2}\hbar$ precisely from a geometric point of view.

Recall that a two-level pure quantum spin state ($|\psi\rangle^{(S=\pm\frac{1}{2}\hbar)}$) can be described by a point on the Bloch sphere $S^2$ (Fig. 1a), and its evolution can be uniquely is represented by the trajectory of the point ($P'$) on the given sphere [1, 2]. Majorana showed in [2] that the problem can be greatly simplified by including more points ($2S$) on the complex projective line $\mathbb{CP}^1$, which can be identified with the Bloch sphere $S^2$ instead of a single point on a higher-dimensional multidimensional geometric structure $(|\psi\rangle_{(2S)}^{(S>j\hbar)}, j > \frac{1}{2})$: $S^D \in \mathbb{R}^{D+1}$, $D \geq 3$. Note that the sphere $S^2$ is equivalent to the factor groups $S^3/S^1$, $SU(2)/U(1)$, $SO(3)/SO(2)$ [3, 4].

A number of papers have been devoted to the study of quantum systems with a high spin using the Majorana representation, in the field of systems with mixed spins [1], quantum entanglement [5], superposition of quantum states of a qutrit – as a quantum cell with three possible states ($|0\rangle$, $|1\rangle$, $|2\rangle$) [6], generalized coherent states [7], adiabatic and superadiabatic processes in three-level systems [8], etc. In particular, in [1] a solution to the problem of describing a two-spin system on the Bloch sphere is proposed, where a practical method is presented for solving the problem of the evolution of a system with mixed spins $(S, \frac{1}{2})$ using $4S + 1$ points on the sphere $S^2$. The work [5] is devoted to a new understanding of the geometry of symmetric entangled states, where an analytical expression is obtained in the case of pure states as a function of the smallest eigenvalue of the Bloch matrix. In [6], the question of the Majorana geometric representation of a qutrit was studied, where a pair of points on the unit sphere $S^2$ represents its quantum states. In this paper [6], a method for the experimental realization of a qutrit system using nuclear magnetic resonance, oriented in a liquid crystal medium is proposed. In [7], based on coherent states, a method was proposed to generalize the Majorana representation (1) for general symmetry. That is, by choosing a generalized coherent state as the reference state, a more general representation of Majorana is given for both finite and infinite systems. Based on this method, the authors of [7] studied the state of compressed vacuum for three different symmetries: Heisenberg-Weyl, SU(2) and SU(1,1) and determined the effect of the compression parameter on the distribution of points on the Bloch sphere.

In this paper, using the Majorana method of representing a quantum pure state with a spin number $S > j\hbar$ ($j \geq \frac{1}{2}$) in terms of $2S$ symmetric states with $S = \frac{1}{2}\hbar$, calculations were made for the correlation functions $F_k(S)$ at $j \geq \frac{1}{2}$. An expression is obtained for the functions $\mathbb{P}_{cor}(F_k(S))$, which describe the probability of matching of oriented spin systems $|\psi\rangle_{(2S)}^{(j\hbar)}$ in the general case: $j \geq \frac{1}{2}$.

**2. Formulation of the problem**

Recall that in [2] for complex numbers $\zeta_s(x_s, y_s)$ the following expression was obtained (Majorana polynomial):

$$a_0\zeta^{2s} + a_1\zeta^{2s-1} + \cdots + a_{2s} = 0, \tag{1}$$

$$a_r = (-1)^r \frac{C_{s-r}}{\sqrt{(2s-r)!\,r!}},$$

where $2S$ complex roots $(\zeta_s)$ under the inverse stereographic projection onto the Bloch sphere (Fig. 1a) form $2S$ points describing the dynamics of unit isovectors with origin at the center of the sphere $S^2$ [1-4]. In this way, the method developed in [2] allows us to construct an expression for the state of the quantum spin system $|\psi\rangle_{(2S)}^{(j\hbar)}$ ($j \geq \frac{1}{2}$) within the three-dimensional space ($S^D \in \mathbb{R}^{D+1}, D < 3$) as a superposition of a system of $2S$ ($S = \pm\frac{1}{2}\hbar$) particles:

$$|\psi\rangle_{(2S)}^{(j\hbar)} = C_s|\psi\rangle_s^{(\hbar/2)} + C_{s-1}|\psi\rangle_{s-1}^{(\hbar/2)} + \cdots + C_{-s}|\psi\rangle_{-s}^{(\hbar/2)}. \tag{2}$$

At the first stage of research in [3], based on the Majorana method, the general form of the system of polynomials (1) was determined:

$$\sum_{r=0}^{2S}(-1)^r \mathbb{N}_r^{(S)} C_{S-r}\zeta^{2S-r} = 0,$$

$$\mathbb{N}_r^{(S)} = \begin{cases} 1, & r = 0 \cup r = 2S \\ \sqrt{\left(\mathbb{N}_{r-1}^{(S-\frac{1}{2})}\right)^2 + \left(\mathbb{N}_r^{(S-\frac{1}{2})}\right)^2}, & 0 < r < 2S \end{cases}.$$

Note that the coefficients $C_S$ ($S = j\hbar, j \geq \frac{1}{2}$) satisfy the normalization condition (Fig. 1)

$$\sum_{r=0}^{2S}|C_r|^2 = 1.$$

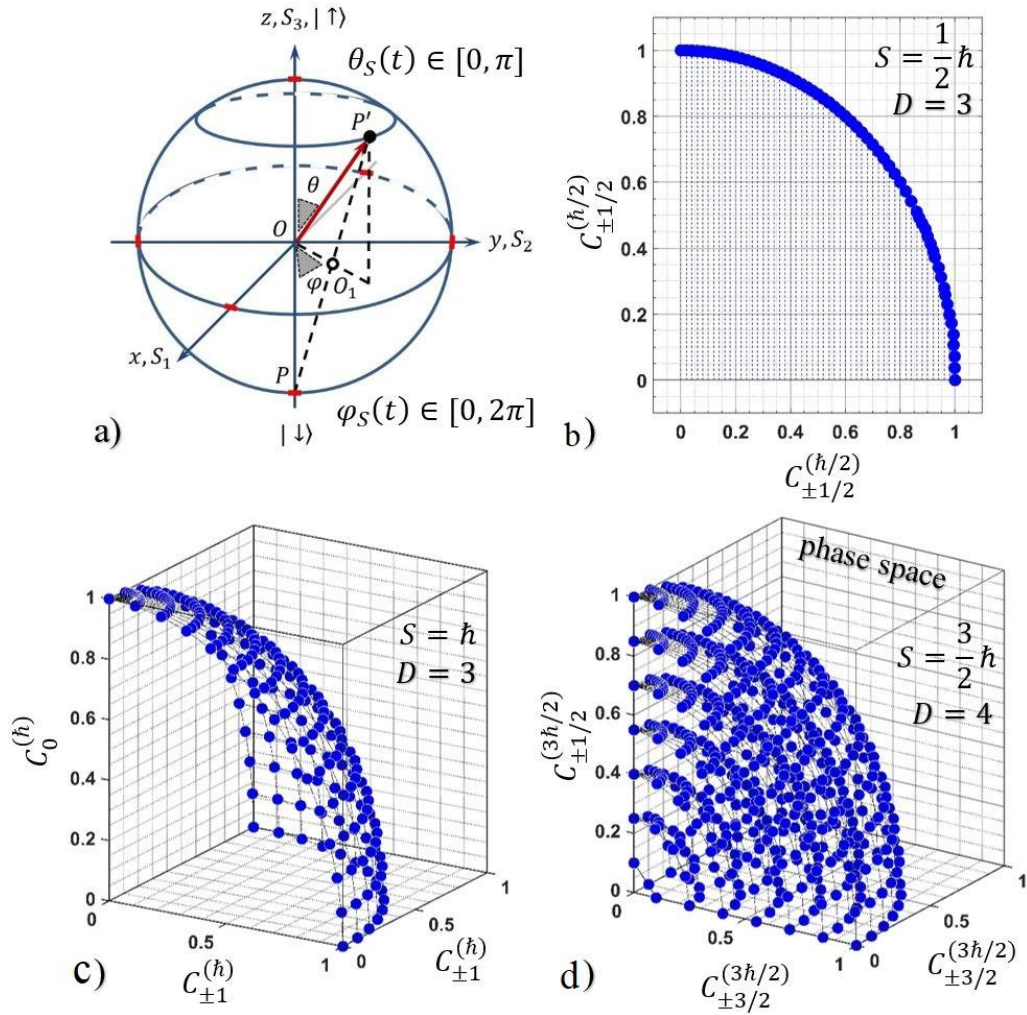

**Figure 1.** Bloch sphere $S^2$ (a). Distribution form of $C_r^{(S\hbar)}$ values at: $S = \frac{1}{2}\hbar$ (b); $S = \hbar$ (c); $S = \frac{3}{2}\hbar$ (d) (in the phase space of the sphere $S^3$).

Range of values $C_S$ ($S > j\hbar, j \geq \frac{1}{2}$) for states with $j = \pm\frac{1}{2}$ (b), $j = \pm 1$ (c) и $j = \pm\frac{3}{2}$ (d – the phase space of the sphere $S^3 \in \mathbb{R}^4$) shown in Fig. 1b, Fig. 1c and Fig. 1d respectively.

The Majorana representation of quantum spin systems by means of ($2S$) representative points ($P'$) becomes particularly simple in the case of oriented states: $\angle POP' = \pi$ (Fig. 1a), where each of the two states is $|\psi\rangle_{(2S)}^{(OP)}, |\psi\rangle_{(2S)}^{(OP')}$ has a spin number $m\hbar$, $m'\hbar$ that is a multiple of $\frac{1}{2}\hbar$. In the general case, the probability of matching these states for $\angle POP' = \alpha$ is defined in [2] in the following form:

$$\mathbb{P}_{cor}(F(\hbar)) = F_1 F_2^{\,2}, \qquad (3),$$

$$F_1(\alpha, j, m, m') = \left(\cos\frac{\alpha}{2}\right)^{4j} (j+m)!\,(j-m)!\,(j+m')!\,(j-m')!,$$

$$F_2(\alpha, r, j, m, m') = \sum_{r=0}^{2j} \frac{(-1)^r \left(\mathrm{tg}\frac{\alpha}{2}\right)^{2r-m+m'}}{r!\,(r-m+m')!\,(j+m-r)!\,(j-m'-r)!}.$$

## 3. Conducted calculations

Let us calculate the values of $\mathbb{P}_{cor}(F_k(S))$ for oriented (Fig. 2) quantum systems with different spin numbers $S = j\hbar$: $j \geq \frac{1}{2}$.

1). $S = \frac{1}{2}\hbar$. Base (fermionic) state with a single point $\zeta_{\frac{1}{2}}$ on the Bloch sphere $S^2$ and two values of the end of the unit isovector $\overrightarrow{OP'}$ in the lower $|\downarrow\rangle$ ($\alpha_1 = 0$) and upper $|\uparrow\rangle$ ($\alpha_2 = 2\pi n$) (more precisely, in diametrically opposite) poles of the sphere $S^2$ (Fig. 2a):

$$\mathbb{P}_{cor}\left(F\left(j=\tfrac{1}{2}, m, m'=\pm\tfrac{1}{2}\right)\right) = \cos^2\frac{\alpha}{2}.$$

2). $S = \hbar$. Two-mode bosonic system with two representative points $\zeta_1$ and spin states – $|\uparrow,\uparrow\rangle, |\uparrow,\downarrow\rangle, |\downarrow,\uparrow\rangle, |\downarrow,\downarrow\rangle$ (Fig. 2b):

$$\mathbb{P}_{cor}(F(j=1, m, m'=\pm 1)) = \cos^4\frac{\alpha}{2},$$

$$\mathbb{P}_{cor}(F(j=1, m, m'=0)) = \cos^2\alpha\,.$$

3). $S = \frac{3}{2}\hbar$. Three-point system $\zeta_{\frac{3}{2}}$ (Fig. 2c):

$$\mathbb{P}_{cor}\left(F\left(j=\tfrac{3}{2}, m, m'=\pm\tfrac{3}{2}\right)\right) = \cos^6\frac{\alpha}{2},$$

$$\mathbb{P}_{cor}\left(F\left(j=\tfrac{3}{2}, m, m'=\pm\tfrac{1}{2}\right)\right) = \cos^2\frac{\alpha}{2}\left[3\cos^2\frac{\alpha}{2} - 2\right]^2.$$

4). General case $(j \geq \frac{1}{2})$: $S = \frac{n}{2}\hbar$: $n = 1, 2, 3, \ldots$ with $2S$ points $\zeta_S$:

$$\mathbb{P}_{cor}(F(S)) = \begin{cases} \cos^{4S}\frac{\alpha}{2} & |m|, |m'| = S_{\frac{n}{2}\hbar} \\ \frac{1}{G}\cos^{4(S-1)}\frac{\alpha}{2}\Psi^2_{(1)}(S) & |m|, |m'| = S_{\frac{n}{2}\hbar-1} \\ \frac{1}{G^2}\cos^{4(S-2)}\frac{\alpha}{2}\Psi^2_{(2)}(S) & |m|, |m'| = S_{\frac{n}{2}\hbar-2} \\ \ldots & \ldots \\ q_k \cos^{4(S-k)}\frac{\alpha}{2}\Psi^2_{(k)}(S) & |m|, |m'| = S_{\frac{n}{2}\hbar-k} \end{cases}, \quad (3^*)$$

$$\Psi^2_k(S) = C_k^{(j)}\cos^k\alpha + C_{k-1}^{(j-1)}\cos^{k-1}\alpha + \cdots + C_1^{(2)}\cos^1\alpha + C_0^{(1)},$$

$$q_k = G^{-k}, \quad k = 0,1,2,3,\ldots, \quad n = 1,2,3,\ldots$$

For the general case $\mathbb{P}_{cor}(F(S))$ ($3^*$) coefficients $C_k^S$ have the following form:

$$\begin{cases} C_0^{(j)} = \{1, \quad j_1 \\ C_1^{(j)} = 2\begin{cases}1 - S, & j_1 \\ S, & j_2 \end{cases} \\ C_2^{(j)} = \begin{cases} 2S^2 - 9S + 8, & j_1 \\ -2(2S^2 - 5S + 2), & j_2 \\ S(2S - 1), & j_3 \end{cases} \\ C_3^{(j)} = 2\begin{cases} -3^{-1}(2S^3 - 21S^2 + 64S - 57), & j_1 \\ 2S^3 - 15S^2 + 31S - 18, & j_2 \\ -2S^3 - 9S^2 + 10S - 3, & j_3 \\ 3^{-1}(2S^3 - 3S^2 + S), & j_4 \end{cases} \\ C_4^{(j)} = 3^{-1}\begin{cases} 2^{-1}(4S^4 - 76S^3 + 491S^2 - 1283S + 1152), & j_1 \\ -2(4S^4 - 60S^3 + 299S^2 - 591S + 396), & j_2 \\ 3(4S^4 - 44S^3 + 155S^2 - 211S + 96), & j_3 \\ -2(4S^4 - 28S^3 + 59S^2 - 47S + 12), & j_4 \\ 2^{-2}(2S - 3)(2S - 2)(2S - 1)S, & j_5 \end{cases} \\ \ldots \end{cases} \quad (4)$$

As mentioned above, in Fig. 2, based on the Majorana representation (1), a graphical description of the oriented states $(\alpha(|\downarrow\rangle) = 0, (|\uparrow\rangle) = 2\pi n, n = 1,2,3, \ldots)$ of quantum spin systems is given (using examples with $S = j\hbar$ ($j = \frac{1}{2}, 1, \frac{3}{2}$)), which are more convenient for theoretical calculations.

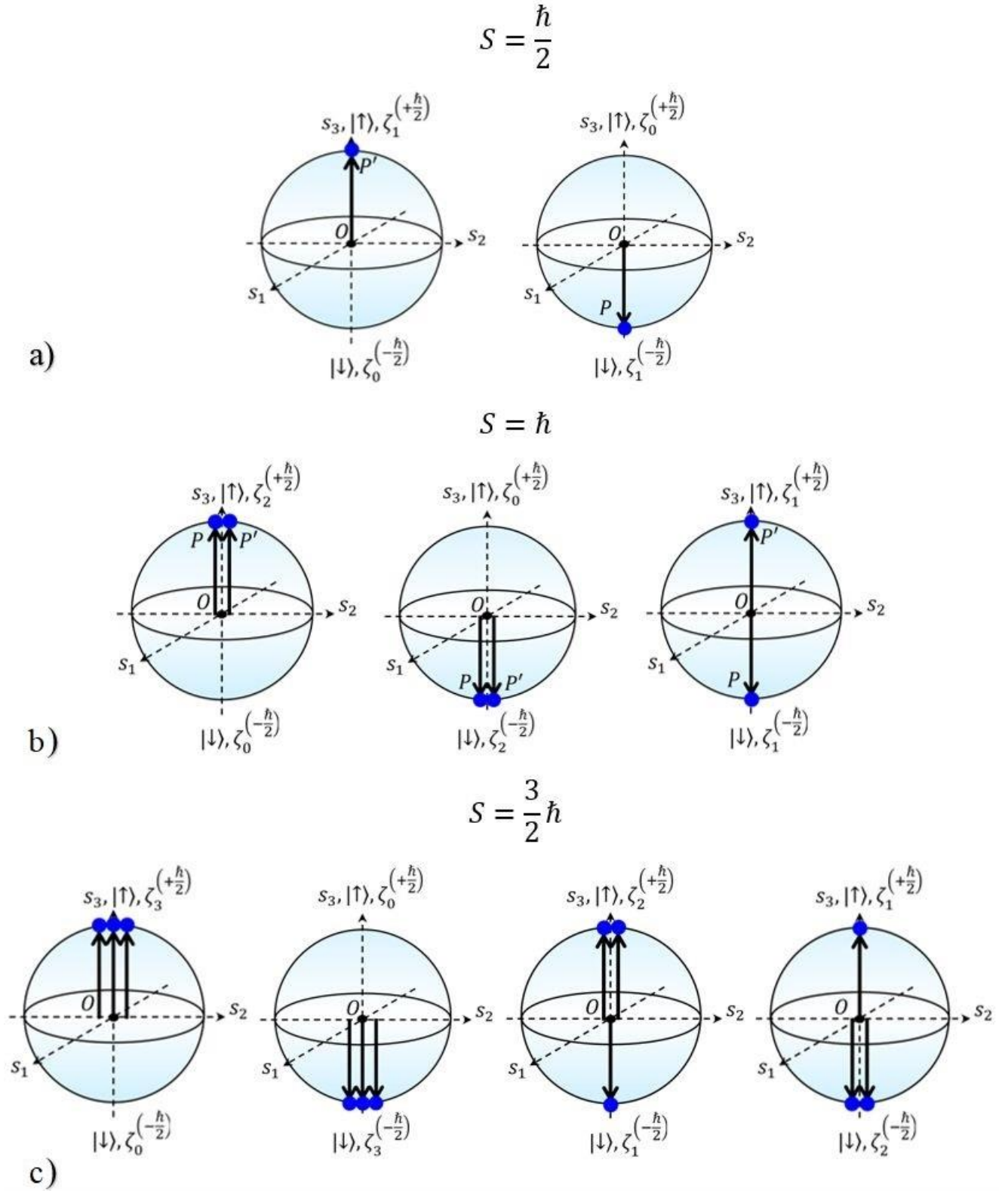

**Figure 2.** The Majorana representation of oriented quantum spin systems as $2S$ symmetric states on the Bloch sphere $S^2$ for: $S = \frac{1}{2}\hbar$ (a); $S = \hbar$ (b); $S = \frac{3}{2}\hbar$ (c).

On Fig. 3 shows the distribution of values $\mathbb{P}_{cor}(F(S))$ (3*) – the probability of matching of oriented spin systems $|\psi\rangle_{(2S)}^{(j\hbar)}$ (2) for different values of $S$. By increasing the values of $\Delta_{j,m,m'} = |j| - |m|$ there is a narrowing of $E(\mathbb{P})$ – the range of values $\mathbb{P}_{cor}(F(S))|_{2\pi k}$ (Fig. 3a – Fig. 3d), which tends to localization by discrete values $2\pi k$ ($k = 0,1,2,3,...$) (Fig. 3e). In this case, some perturbations $E(\mathbb{P})$ (Fig. 3b – Fig. 3e), obviously, are a consequence of the influence of the polynomial function $\Psi_k^2(S)$ (3*), (4)

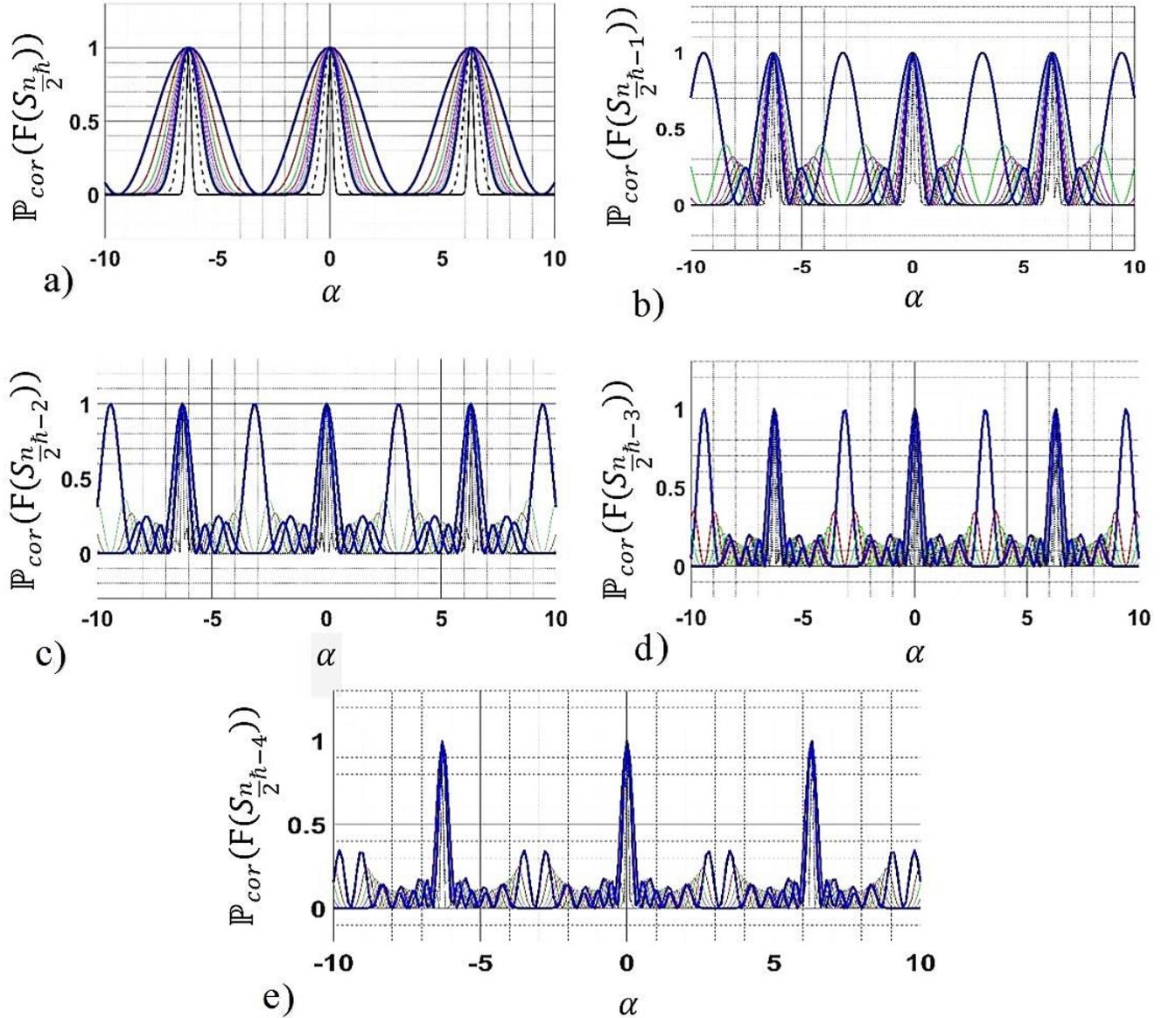

**Figure. 3.** Distribution of $\mathbb{P}_{cor}(F(S))$ (3*) values for different $S$ values: $\frac{n}{2}\hbar$ (a); $\frac{n}{2}\hbar - 1$ (b); $\frac{n}{2}\hbar - 2$ (c); $\frac{n}{2}\hbar - 3$ (d); $\frac{n}{2}\hbar - 4$ (e).

## 4. Conclusions

The Majorana representation is an efficient, intuitive way to explore the fundamental properties of multidimensional quantum states. This method also has an extremely wide range of practical and promising applications, ranging from quantum optics, which makes it possible to design arbitrary quantum states from coherent ones, to high-precision generation of distributed multi-qubit entanglement for large-scale quantum communication and computing networks [8, 9]. This paper presents the results of calculations using the Majorana method to describe the correlation functions, as well as the matching probability of oriented quantum systems $|\psi\rangle_{(2S)}^{(j\hbar)}$ (2) for different values of the spin number $S = j\hbar$. In particular, the general form of the matching probability distribution function for oriented spin systems $\mathbb{P}_{cor}(F(S))$ is determined.

**Funding**

The work was carried out as part of the implementation of the research plan of the Department of Nanomaterials and Nanotechnologies of the S.U.Umarov Physical–Technical Institute of the National Academy of Sciences of Tajikistan, grant No. 0119TJ00994.